\documentclass[aps,prl,twocolumn,showpacs,psfig,superscriptaddress,longbibliography]{revtex4-1}

\usepackage{textcomp}
\usepackage{times}
\usepackage{graphicx}
\usepackage{float}
\usepackage{latexsym,amsmath,amssymb,bm,euscript}
\usepackage{color}
\usepackage{subfigure}
\usepackage{epstopdf}
\usepackage[colorlinks=true,linkcolor=blue,citecolor=blue]{hyperref}
\usepackage{hyperref}
\usepackage{soul}
\usepackage[normalem]{ulem}
\usepackage{mathrsfs}
\usepackage{amsmath}
\usepackage{lettrine}
\usepackage{xspace}
\usepackage{textcomp}

\begin{document}

\title{Two Plaquette-Singlet Phases and Emergent SO(5) Deconfined Quantum Criticality in  SrCu$_2$(BO$_3$)$_2$}

\author{Yi Cui}
\thanks{These authors contributed equally to this study.}
\affiliation{School of Physics and Beijing Key Laboratory of
Opto-electronic Functional Materials $\&$ Micro-nano Devices, Renmin
University of China, Beijing, 100872, China}
\affiliation{Key Laboratory of Quantum State Construction and
Manipulation (Ministry of Education), Renmin University
of China, Beijing, 100872, China}

\author{Kefan Du}
\thanks{These authors contributed equally to this study.}
\affiliation{School of Physics and Beijing Key Laboratory of
Opto-electronic Functional Materials $\&$ Micro-nano Devices, Renmin
University of China, Beijing, 100872, China}

\author{Zhanlong Wu}
\thanks{These authors contributed equally to this study.}
\affiliation{School of Physics and Beijing Key Laboratory of
Opto-electronic Functional Materials $\&$ Micro-nano Devices, Renmin
University of China, Beijing, 100872, China}

\author{Shuo Li}
\affiliation{Beijing National Laboratory for Condensed Matter Physics and
Institute of Physics, Chinese Academy of Sciences, Beijing, 100190, China}

\author{Pengtao Yang}
\affiliation{Beijing National Laboratory for Condensed Matter Physics and
Institute of Physics, Chinese Academy of Sciences, Beijing, 100190, China}

\author{Ying Chen}
\affiliation{School of Physics and Beijing Key Laboratory of
Opto-electronic Functional Materials $\&$ Micro-nano Devices, Renmin
University of China, Beijing, 100872, China}

\author{Xiaoyu Xu}
\affiliation{School of Physics and Beijing Key Laboratory of
Opto-electronic Functional Materials $\&$ Micro-nano Devices, Renmin
University of China, Beijing, 100872, China}

\author{Hongyu Chen}
\affiliation{School of Physics and Beijing Key Laboratory of
Opto-electronic Functional Materials $\&$ Micro-nano Devices, Renmin
University of China, Beijing, 100872, China}

\author{Chengchen Li}
\affiliation{School of Physics and Beijing Key Laboratory of
Opto-electronic Functional Materials $\&$ Micro-nano Devices, Renmin
University of China, Beijing, 100872, China}

\author{Juanjuan Liu}
\affiliation{School of Physics and Beijing Key Laboratory of
Opto-electronic Functional Materials $\&$ Micro-nano Devices, Renmin
University of China, Beijing, 100872, China}
\affiliation{Key Laboratory of Quantum State Construction and
Manipulation (Ministry of Education), Renmin University
of China, Beijing, 100872, China}

\author{Bosen Wang}
\thanks{These authors contributed equally to this study.}
\affiliation{Beijing National Laboratory for Condensed Matter Physics and
Institute of Physics, Chinese Academy of Sciences, Beijing, 100190, China}

\author{Wenshan Hong}
\affiliation{Beijing National Laboratory for Condensed Matter Physics and Institute of Physics,
Chinese Academy of Sciences, Beijing, 100190, China}

\author{Shiliang Li}
\affiliation{Beijing National Laboratory for Condensed Matter Physics and
Institute of Physics, Chinese Academy of Sciences, Beijing, 100190, China}
\affiliation{School of Physical Sciences, University of Chinese Academy of Sciences, Beijing, 100190, China}

\author{Zhiyuan Xie}
\affiliation{School of Physics and Beijing Key Laboratory of
Opto-electronic Functional Materials $\&$ Micro-nano Devices, Renmin
University of China, Beijing, 100872, China}
\affiliation{Key Laboratory of Quantum State Construction and
Manipulation (Ministry of Education), Renmin University
of China, Beijing, 100872, China}

\author{Jinguang Cheng}
\affiliation{Beijing National Laboratory for Condensed Matter Physics and
Institute of Physics, Chinese Academy of Sciences, Beijing, 100190, China}
\affiliation{School of Physical Sciences, University of Chinese Academy of Sciences, Beijing, 100190, China}

\author{Bruce Normand}
\affiliation{PSI Center for Scientific Computing, Theory and Data, CH-5232 Villigen-PSI, Switzerland}

\author{Rong Yu}
\affiliation{School of Physics and Beijing Key Laboratory of
Opto-electronic Functional Materials $\&$ Micro-nano Devices, Renmin
University of China, Beijing, 100872, China}
\affiliation{Key Laboratory of Quantum State Construction and
Manipulation (Ministry of Education), Renmin University
of China, Beijing, 100872, China}

\author{Weiqiang Yu}
\email{wqyu\_phy@ruc.edu.cn}
\affiliation{School of Physics and Beijing Key Laboratory of
Opto-electronic Functional Materials $\&$ Micro-nano Devices, Renmin
University of China, Beijing, 100872, China}
\affiliation{Key Laboratory of Quantum State Construction and
Manipulation (Ministry of Education), Renmin University
of China, Beijing, 100872, China}


\begin{abstract}

The deconfined quantum critical point (DQCP) has become a central open concept in the physics of quantum matter, and its proposed  presence in the Shastry-Sutherland model was followed by the experimental observation of at least a minimal DQC scenario induced by an applied magnetic field in SrCu$_2$(BO$_3$)$_2$. However, the nature of the plaquette-singlet phase in SrCu$_2$(BO$_3$)$_2$ remains unresolved, and with it the identification of the DQCP symmetry from among several theoretical scenarios. Here we perform detailed high-pressure $^{11}$B NMR studies to reveal the presence of both the full-plaquette (FP) and empty-plaquette (EP) phases in SrCu$_2$(BO$_3$)$_2$, phase-separated at a first-order, pressure-driven transition with a volume-fraction effect. The field-driven transition from the EP to the antiferromagnetic (AFM) phase complements our previous observations of the FP--AFM transition, with both showing deconfined quantum criticality, while the scaling of the spin-lattice relaxation rate near the EP--AFM transition, $1/T_1 \propto T^{0.6}$, suggests a DQCP governed by a different universality class. We discuss possible extensions to the  Shastry-Sutherland model that account for these pressure and field effects. The expanded phase space we discover mandates an SO(5) DQCP symmetry, and hence our results take an important step towards a complete understanding of deconfined quantum criticality in SrCu$_2$(BO$_3$)$_2$.
\end{abstract}

\maketitle

\begin{figure*}[t]
\centering
\includegraphics[width=17.8cm]{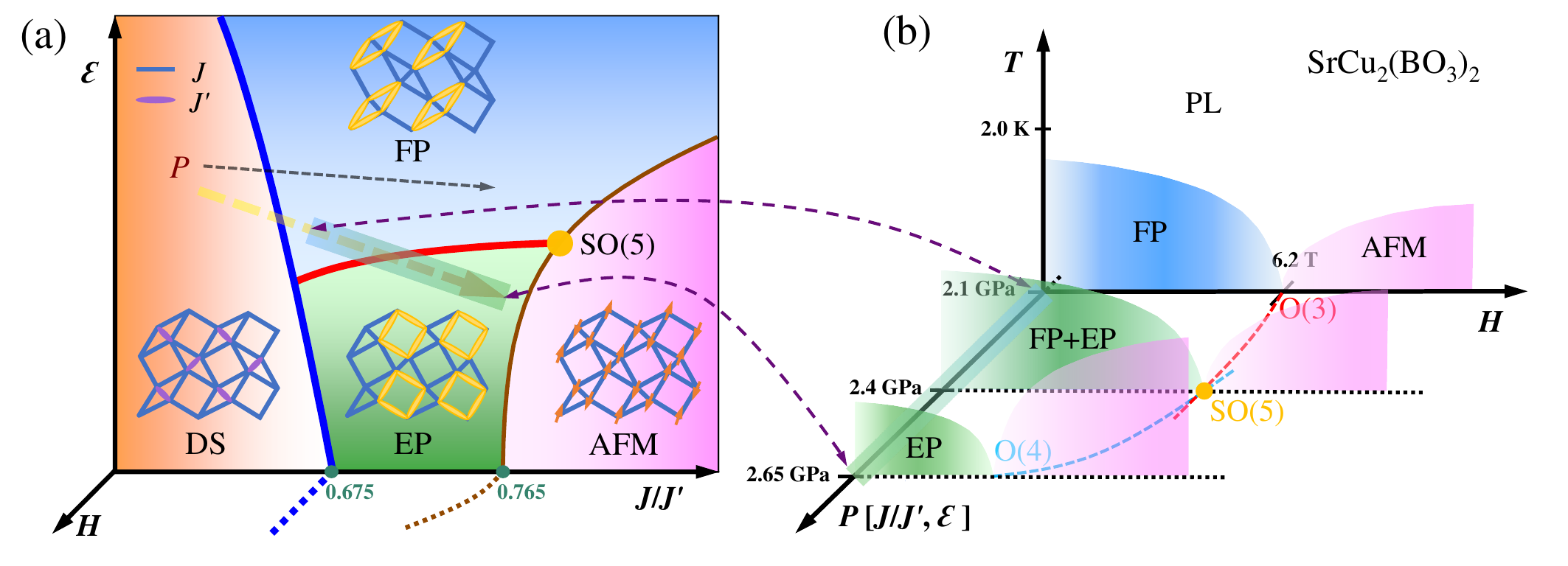}
\caption{\label{pd2d}
{\bf Phase diagrams of the SSM and SrCu$_2$(BO$_3$)$_2$ under an applied hydrostatic pressure and magnetic field.} (a) Ground-state phase diagram based on an extended Shastry-Sutherland model at zero field and temperature. $J'$ and $J$ label the intra- and interdimer magnetic interactions of the Shastry-Sutherland model and $\varepsilon$ is an additional parameter that governs the coupling of the spin and lattice sectors (see text). The thick dotted line represents the pressure trajectory consistent with the current experiments and the thin dotted line represents that consistent with the experiment of Ref.~\cite{Cui_Science_2023}. Phase boundaries drawn with thick lines (blue and red) represent first-order quantum phase transitions and that drawn with a thin line (brown) a second-order transition. An emergent SO(5) symmetry is proposed at the multicritical point. (b) Three-dimensional phase diagram of the plaquette-singlet phases in the space of pressure, field, and temperature. The field favors the AFM phase and hence induces both FP--AFM and EP--AFM transitions, with respective enhanced symmetries O(3) and O(4). Emergent SO(5) deconfined quantum criticality is manifest where the two lines cross. Data representing 2.1~GPa are taken from Ref.~\cite{Cui_Science_2023}. }
\end{figure*}

{\it Introduction---}Magnetic frustration induces strong quantum fluctuations, making low-dimensional, low-spin quantum magnets an excellent platform in the search for emergent quantum phases and quantum phase transitions (QPTs). A notable example is the Shastry-Sutherland model (SSM)~\cite{Shastry_1981} of $S = 1/2$ spins on the orthogonal dimer lattice [insets, Fig.~\ref{pd2d}(a)], with competing antiferromagnetic (AFM) Heisenberg interactions $J'$ on the dimers and $J$ between them. Increasing the ratio $\alpha = J/J'$ drives QPTs from a dimer-singlet (DS) phase to a plaquette-singlet (PS) phase~\cite{Koga_PRL_2000, Moliner_PRB_2011}, and then to an ordered AFM phase~\cite{Albrecht_EPL_1996, Miyahara_PRL_1999, Hartmann_PRL_2000, Koga_PRL_2000, Lauchli_PRB_2002, Corboz_PRB_2013}. A key theoretical prediction in this model~\cite{Lee_PRX_2019} is that the PS--AFM phase transition is a deconfined quantum critical point (DQCP)~\cite{Senthil_Science_2004}, a continuous order-to-order QPT beyond the Landau paradigm with attendant enhanced symmetries and fractional excitations~\cite{Senthil_Science_2004, Sandvik_PRL_2007, Nahum_PRL_2015, Shao_Science_2016, Ma_PRB_2018, Zhao_NP_2019, Serna_PRB_2019}. Nevertheless, detailed numerical calculations on extended SSMs remain divided on supporting the DQCP scenario~\cite{Guzc_PRL_2024} or challenging it by reporting a (weakly) first-order PS--AFM transition~\cite{Li_PRL_2023, Xi_PRB_2023}. The realization of the SSM in material form would therefore be of great interest, offering the possibility that experiment could verify DQCP phenomena such as spinons at the critical point.

While studies of deconfined quantum criticality have extended to diverse physical systems~\cite{Nahum_PRX_2015, Wang_PRX_2017, Mengzy_PRX_2017, Zhao_PRL_2020, LiuZH_PRL_2022, ShuYR_PRL_2022}, the closest to realization remains the SSM, due to the remarkable properties of the compound SrCu$_2$(BO$_3$)$_2$~\cite{kageyama_JPSJ_1998, Kageyama_PRL_1999}. At ambient pressure,
SrCu$_2$(BO$_3$)$_2$ presents as an orthogonal dimer lattice of Cu$^{2+}$ ions in the DS phase, with the best estimates for the interactions being $J' \simeq 81.5$~K and $J \simeq 51.3$~K ($\alpha = 0.63$)~\cite{Shi_NC_2022}, and further interactions being below 3\% of $J'$~\cite{Cepas_PRL_2001,Fogh_PRL_2024}. $\alpha$ grows with increasing pressure, making SrCu$_2$(BO$_3$)$_2$ a near-ideal candidate for exploring the phase diagram of the SSM~\cite{Zayed_NP_2017,Guo_PRL_2020,Larrea_Nature_2021}. While zero-field specific-heat measurements suggest that the pressure-induced PS--AFM transition is first-order~\cite{Guo_2023}, high-pressure NMR studies have indicated a field-induced PS--AFM transition with a proximate DQCP~\cite{Cui_Science_2023}. However, SrCu$_2$(BO$_3$)$_2$ differs from the SSM in that the PS phase of the SSM has no $J'$ dimer bond inside the ``empty plaquette'' (EP) formed from four $J$ bonds making a local plaquette singlet [Fig.~\ref{pd2d}(a)]~\cite{Koga_PRL_2000, Moliner_PRB_2011,Corboz_PRB_2013}. By contrast, inelastic neutron scattering (INS)~\cite{Zayed_NP_2017} and NMR~\cite{Cui_Science_2023} studies on SrCu$_2$(BO$_3$)$_2$ seem to favor a PS phase with the plaquettes formed around the $J'$ bonds, known as ``full plaquette'' (FP) [Fig.~\ref{pd2d}(a)]~\cite{Waki_JPSJ_2007, Moliner_PRB_2011}. The consequences of this PS-phase competition for the DQCP in SrCu$_2$(BO$_3$)$_2$ have yet to be understood.

Here we show that both the FP and the EP phases are present in SrCu$_2$(BO$_3$)$_2$. By detailed high-pressure $^{11}$B NMR studies with careful spectral analysis, we identify a pressure-induced first-order phase transition from the FP to the EP, with evidence of phase separation and a change of volume fraction. The macroscopic phase coexistence of EP with FP is reflected in a splitting of the NMR line, with the high-frequency peak corresponding to the FP phase studied in Ref.~\cite{Cui_Science_2023} and the low-frequency one, not previously detected, corresponding to the EP. We find a field-induced EP--AFM transition at which the energy scales of both phases vanish around 5.5~T at 2.4~GPa, indicating the proximate deconfined quantum critical behavior reported at the field-induced FP--AFM transition~\cite{Cui_Science_2023}, but with the scaling of the spin-lattice relaxation rate in the critical regime, $1/T_1 \propto T^{0.6}$, characteristic of a different universality class. These discoveries add one dimension to the phase space of SrCu$_2$(BO$_3$)$_2$ and point to the realization of a higher-symmetry DQCP than the minimal model used previously to interpret the PS--AFM transition.

\begin{figure}[t]
\centering
\includegraphics[width=8.5cm]{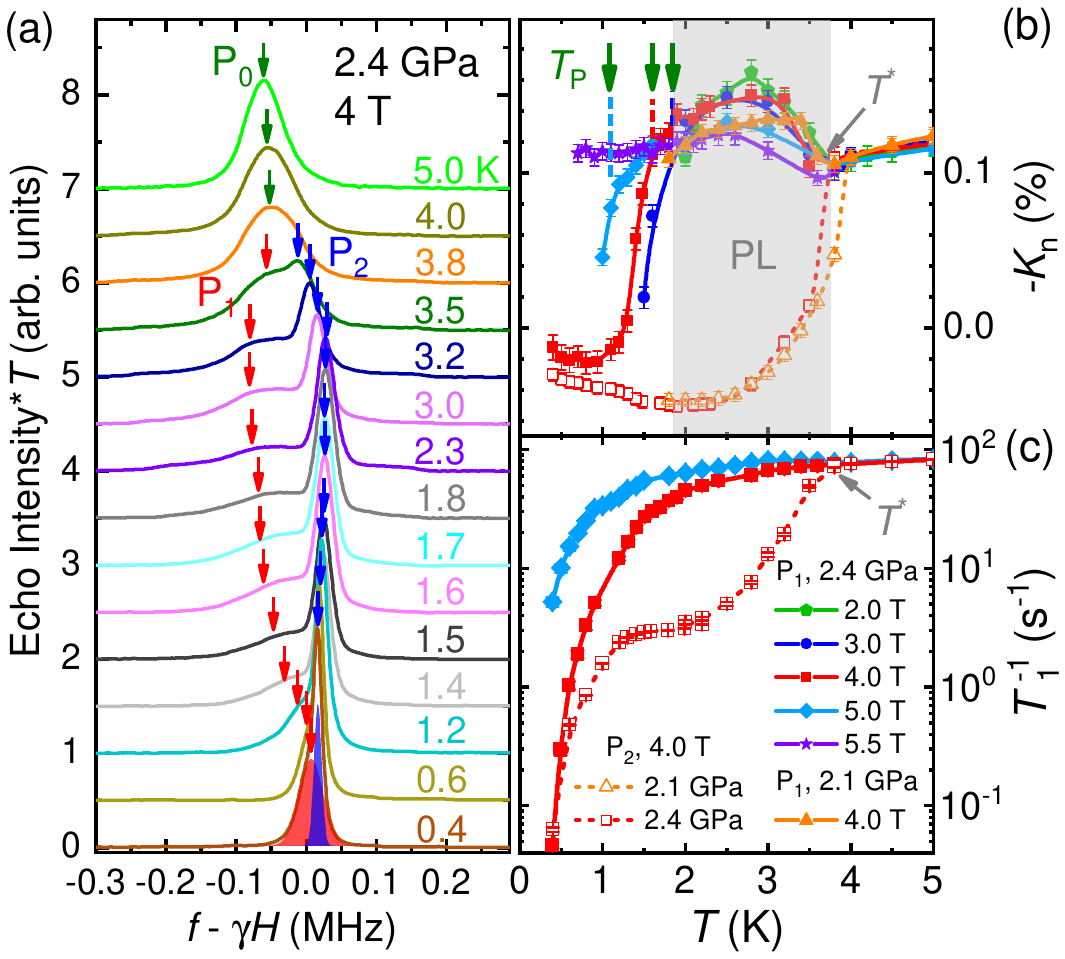}
\caption{\label{spec}
{\bf $^{11}$B NMR spectra taken at 2.4~GPa.} (a) Spectra measured at 4~T for a sequence of decreasing temperatures (data offset vertically for clarity). Peaks labelled as P$_0$, P$_1$, and P$_2$ indicate the positions of the dominant features. An example of double-Gaussian fitting, with the resulting red and blue lineshapes, is shown for the spectrum at 0.4~K. (b) Knight shifts, $K_{\rm n} (T)$, for the different peaks, measured at five magnetic fields. The PS ordering temperature, $T_{\rm P}$, is identified by the sharp drop in $-K_{\rm n}$ for the P$_1$ peak upon cooling. $T^*$ marks the onset temperature of the PL phase (shaded area). (c) $1/T_1 (T)$ measured at the positions of the different peaks.}
\end{figure}

\begin{figure}[t]
\centering
\includegraphics[width=8.5cm]{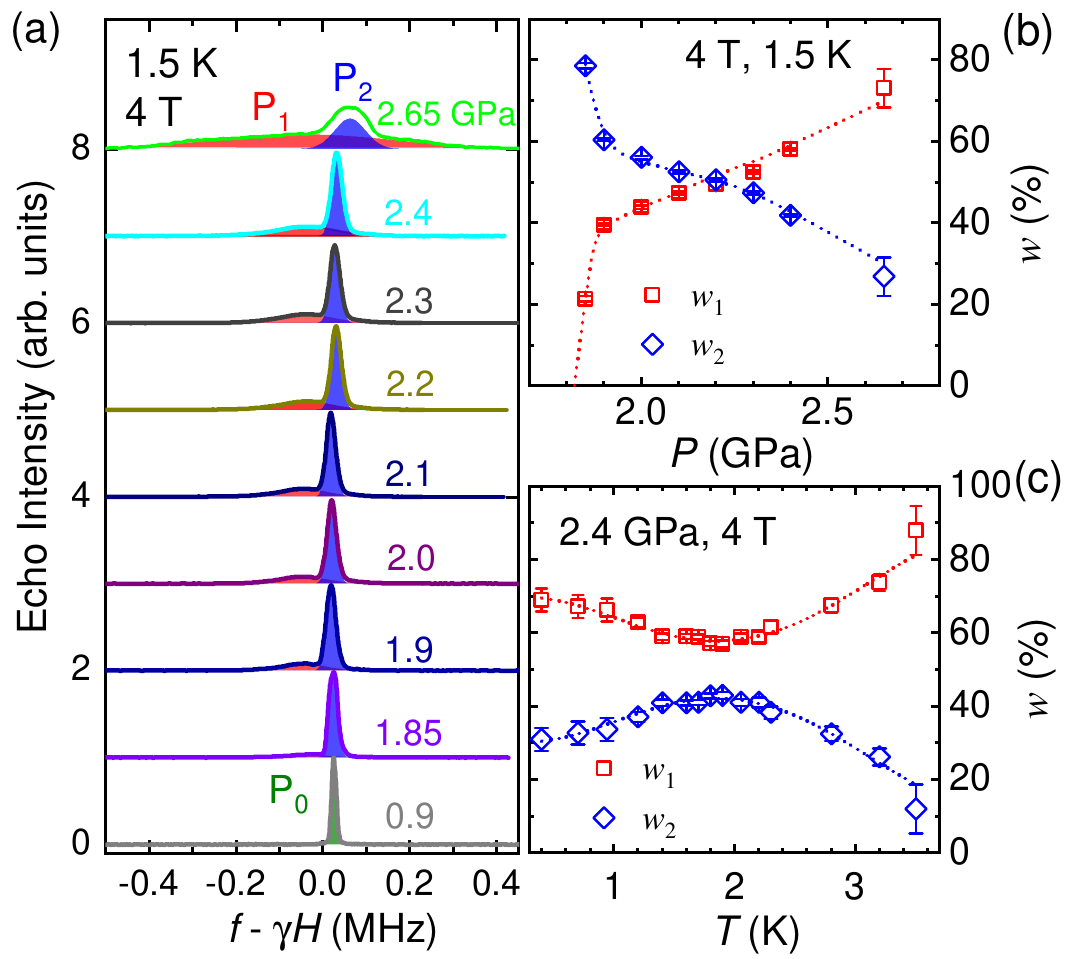}
\caption{\label{ratio} {\bf Volume fractions of two different PS phases.} (a) NMR spectra measured at 1.5~K and 4~T under increasing pressure (data offset vertically for clarity). The green, blue, and red shaded areas represent the results of single- or double-Gaussian fits, corresponding respectively to the peak positions P$_0$, P$_1$, and P$_2$. (b) Volume fractions $w_1$ and $w_2$ measured at 4~T and 1.5~K and shown as functions of pressure. (c) Temperature-dependence of $w_1$ and $w_2$ measured at 2.4~GPa and 4~T.}
\end{figure}

{\it Phase separation of distinct PS phases---}$^{11}$B NMR measurements were conducted on a single crystal with the magnetic field applied along the crystalline $c$-axis. A NiCrAl piston cell was used to achieve pressures up to 2.4~GPa, and a cubic anvil cell was employed to reach 2.65~GPa. The center NMR transition lines (spin $I = -1/2$ to $1/2$) of the $^{11}$B spectra at 2.4~GPa and 4~T, measured from 5~K down to 0.4~K, are shown in Fig.~\ref{spec}(a). At temperatures down to 3.8~K we observe a single resonance line (labelled P$_0$), but further cooling causes the resonance to split into two distinct peaks, denoted P$_1$ and P$_2$, which we separated by double-Gaussian fits to all spectra, including in our subsequent analysis at all measured recovery times (which satisfied $t > 10 \, T_1$). P$_2$ is a narrow, high-frequency peak whose position shows a monotonic increase with decreasing temperature, as reported previously for the FP phase~\cite{Waki_JPSJ_2007, Moliner_PRB_2011, Cui_Science_2023}. By contrast, P$_1$ is a broad peak at low frequencies, whose position shifts first to lower and then to higher frequencies as the temperature decreases.

\begin{figure}[t]
\centering
\includegraphics[width=8.2cm]{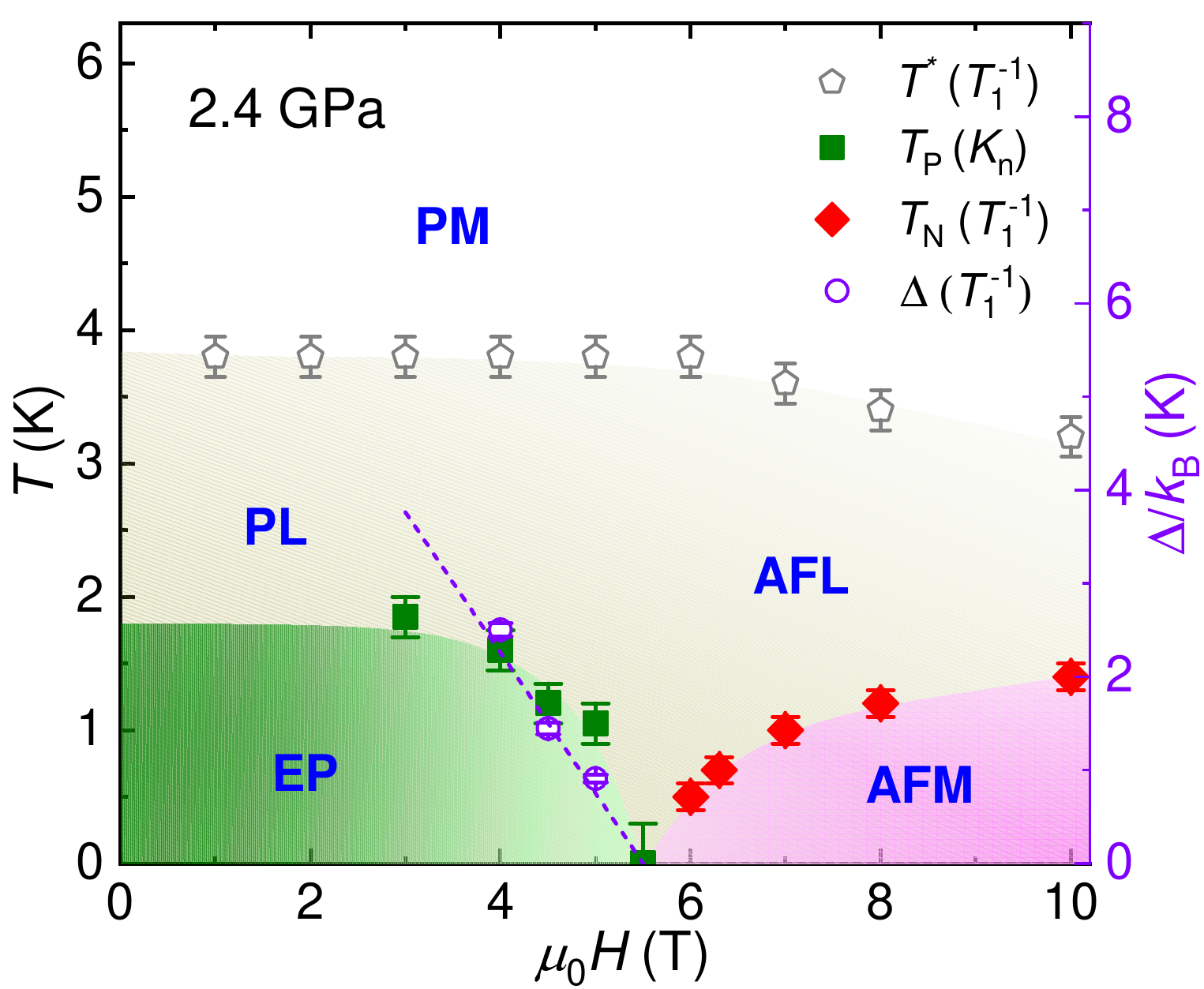}
\caption{\label{pd} {\bf Field-temperature phase diagram deduced from P$_1$ at 2.4~GPa.} $T^*$ marks the onset temperature of magnetic correlations in the paramagnetic (PM) state. These establish a plaquette-liquid (PL) regime at low field and an antiferromagnetic liquid (AFL) regime at high field. $T_{\rm N}$ is the N\'eel temperature of the ordered AFM phase and $T_{\rm P}$ the ordering transition of the EP phase. $\Delta$ (right axis) is the energy gap determined in the EP phase. The dotted line shows a linear fit of the form $\Delta(H) = g \mu_B (H_{\rm C} - H)$. }
\end{figure}

The Knight shifts, $K_{\rm n}(T)$, for both P$_1$ and P$_2$ were calculated using the negative hyperfine coupling constant $A_{\rm hf} \simeq - 0.259$~T/$\mu_B$ \cite{kodama_JPCM_2002, Cui_Science_2023}, and are shown at different magnetic fields in Fig.~\ref{spec}(b). At all fields, $-K_{\rm n}$ for P$_2$ exhibits a sharp downturn around 3.5~K, which was identified in Ref.~\cite{Cui_Science_2023} as the onset of spin correlations on the plaquette; this temperature,  $T^*$, is not a phase transition but marks the emergence of a fluctuating PS phase termed the plaquette liquid (PL). By contrast, $-K_{\rm n}$ for P$_1$ exhibits an upturn at $T^*$, indicating a quite different type of plaquette fluctuations. It then peaks between 3 and 2~K before dropping sharply at a field-dependent temperature below 2~K that we identify as $T_{\rm P}$, the ordering temperature [indicated by down-pointing arrows in Fig.~\ref{spec}(b)] of a single, gapped PS phase, which is also visible as a reduction in the uniform magnetic susceptibility. We remark here that both the FP and EP phases have a double degeneracy [Fig.~\ref{pd2d}(a)], which is broken below the Z$_2$ (Ising) transition at $T_{\rm P}$, whereas the PL phase is a fluctuation between the two. The values of $T_{\rm P}$ and $T^*$ extracted for all our measurement fields are shown in the phase diagram of Fig.~\ref{pd}.

\begin{figure}[t]
\centering
\includegraphics[width=8.5cm]{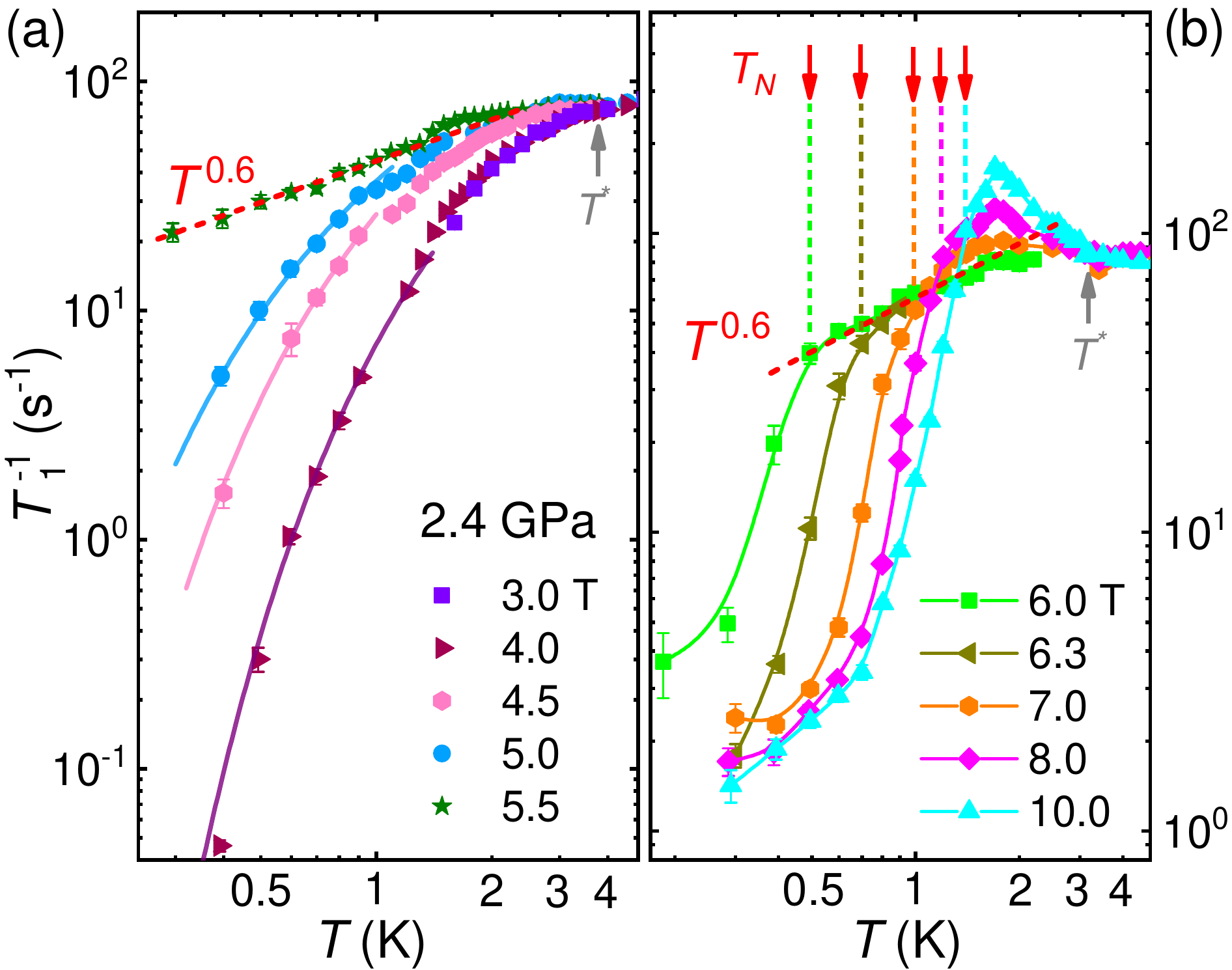}
\caption{\label{relax} {\bf Spin-lattice relaxation rates measured on P$_1$ at 2.4 GPa.} (a) $1/T_1 (T)$ measured at five low magnetic fields. Solid lines represent fits to the thermal activation function, given by $1/T_1 \propto T^{0.6} e^{-\Delta/k_{\rm B}T}$. (b) $1/T_1 (T)$ measured at five high magnetic fields. The N\'{e}el temperature, $T_{\rm N}$, is indicated by the sharp drop in $1/T_1$. Dotted lines in both panels serve as guides to the eye, representing the scaling behavior $1/T_1 \propto T^{0.6}$ at 5.5~T and at 6~T.}
\end{figure}

The spin-lattice relaxation rates, $1/T_1$, measured at peaks P$_1$ and P$_2$ are compared in Fig.~\ref{spec}(c). For P$_2$, $1/T_1(T)$ shows the same sharp drop below $T^*$ as $K_{\rm n}(T)$, followed by a plateau below 2.5~K before $T_{\rm P}$, invisible in $K_{\rm n}(T)$, can be identified from the further drop off this plateau~\cite{Cui_Science_2023}. For P$_1$, $1/T_1$ shows a small increase at $T < T^*$, before a steady drop with decreasing temperature for all fields that is insensitive to $T_{\rm P}$, and which we investigate further in Fig.~\ref{relax}. Thus the spin dynamics reflected in P$_1$ and P$_2$ are distinctly different, with the latter indicating the opening of a spin gap at $T^*$ whereas the former shows enhanced spin fluctuations in the PL regime, suggesting gapless spin excitations. The P$_1$ and P$_2$ signals become similar again only well below $T_{\rm P}$, where the $K_{\rm n}$ values approach zero as expected for a gapped phase.

{\it First-order FP--EP transition---} In Fig.~\ref{ratio}(a) we investigate the evolution of peaks P$_1$ and P$_2$ with pressure, measured at a constant field and temperature $T < T_{\rm P}$. Only a single peak is observed at 0.9~GPa, but a double-peak structure emerges at and above 1.85~GPa. From the double-Gaussian fits represented by the red and blue peaks, it is clear that the relative spectral weight of P$_1$ increases with pressure. We then calculate the volume fraction of the P$_1$ phase, $w_1$, from its spectral weight relative to the total spectral weight, and similarly for P$_2$ ($w_2 = 1 - w_1$). The evolution of $w_1$ and $w_2$ with pressure across the PS phase is shown in Fig.~\ref{ratio}(b).  $w_1$ exhibits a rapid increase between 1.8~GPa and 1.9~GPa, just after the system leaves the DS phase, followed by a more gradual and largely linear to rise 70\% at 2.65~GPa, indicating that the P$_1$ phase is increasingly favored at higher pressures. We note here that the spin-spin relaxation rates, $1/T_2$, were found to be different on P$_1$ and P$_2$ (data not shown), and hence all data for these lines were fully calibrated to account for $T_2$ effects. At a fixed pressure of 2.4~GPa and a field of 4~T, where $T_{\rm P} = 1.6$~K, spectral measurements conducted at various temperatures show a slight decrease in $w_1$ up to 1.8~K, as shown in Fig.~\ref{ratio}(c), followed by an increase towards 100\% across the PL phase. We also confirm for each measurement pressure that the variation in $w_1$ with field is negligible up to 5.5~T (data not shown).

With their quite different spin dynamics, it is clear that P$_1$ and P$_2$ are distinct phases, and their volume-fraction effect makes clear a macroscopic phase coexistence, as expected at a first-order phase transition. Turning to theory for insight, tensor-network calculations for the EP phase of the SSM find a prominent peak in the specific heat ($C_{\rm m}/T$) at $T \approx 0.04 J'$, indicating strong PL fluctuations~\cite{Li_PRL_2023}. Taking $J' \approx 50$~K at these pressures, this is consistent with the peak around 2.5~K in our $K_{\rm n}$ measurements for P$_1$ [Fig.~\ref{spec}(b)]. We therefore propose that the phase associated with P$_1$ is the EP phase of the SSM, as indicated in Fig.~\ref{pd2d}(a). In contrast, we associate the gapped behavior of P$_2$ below $T^*$ with the FP phase~\cite{Cui_Science_2023}. Here the result that $\langle {\vec S}_1 \cdot {\vec S}_2 \rangle \approx 0$ on the intradimer bond in the EP phase \cite{Larrea_Nature_2021} suggests a loss of influence from the DS gap, whereas one anticipates [insets, Fig.~\ref{pd2d}(a)] that this bond will retain significant negative spin correlations, and hence a gap effect, in the FP phase. The presence of an FP component indicates that an extended SSM is required, such as one with a lattice distortion favoring one of the two FP states~\cite{Boos_PRB_2019}.

As noted above, the large increase of $w_1$ with pressure and the phase coexistence imply a first-order phase transition from FP to EP. This is illustrated in Fig.~\ref{pd2d}(a), where we propose that the extra parameter, $\varepsilon$, is a relative spin-lattice coupling. While a static lattice distortion of the type considered in Ref.~\cite{Boos_PRB_2019} is excluded by the absence of NMR line-splitting in the DS phase, a spontaneous lattice distortion analogous to a spin-Peierls effect is not. At large $\varepsilon$, the gain in magnetic energy from a robust PS exceeds the lattice energy cost, and a distortion selects one of the two FP states. At small $\varepsilon$, the lattice is not involved and the EP phase of the SSM emerges. A hydrostatic pressure changes not only $J/J'$ but also $\varepsilon$, and we suggest the trajectory shown in Fig.~\ref{pd2d}(a) on the basis that higher pressures make any lattice distortions more costly. In this way the pressure drives an FP--EP transition, whose nature is manifestly first-order.

We remark here that the INS structure factor measured in SrCu$_2$(BO$_3$)$_2$ at a pressure just above the DS--PS boundary was found to be more consistent with the FP phase~\cite{Zayed_NP_2017}. By contrast, $C_{\rm m}/T$ data show a broad peak between 2.0~K and 3.2~K~\cite{Guo_2023, Larrea_Nature_2021}, which aligns with our $K_{\rm n}$ and $1/T_1$ data for the PL state above the EP. Given that the EP and FP phases are nearly degenerate~\cite{Moliner_PRB_2011}, these discrepancies may arise due to the differing sensitivities of the two phases to different probes, as well as to the degree of hydrostaticity actually achieved in each  pressure cell and experiment. On this topic we remark that the N\'eel temperature we observe in the AFM phase (Fig.~\ref{pd}) is consistently over 50\% higher than in Ref.~\cite{Cui_Science_2023}, one interpretation for which may be that the high-pressure phase is more homogeneous in our current experiment.

{\it Field-induced EP--AFM transition---}The temperature-dependence of $1/T_1$ was measured on P$_1$ over a wide range of magnetic fields at 2.4~GPa, as shown in Fig.~\ref{relax}. At low temperatures, $1/T_1$ increases with field strength from 3~T to 5.5~T but decreases from 6~T to 10~T, indicating a transition from the PS (EP) phase to the AFM phase~\cite{Cui_Science_2023}. At 5.5~T, the absence of a downturn in $1/T_1$ down to 0.3~K suggests that the EP--AFM transition field is close to this value. In the PL phase, $1/T_1$ follows a power-law dependence, $1/T_1 \propto T^{0.6}$, over the temperature range from 0.3~K to 1.5~K at 5.5~T [Fig.~\ref{relax}(a)] and from 0.5~K to 2~K at 6.0~T [Fig.~\ref{relax}(b)].

For fields below 5~T, $1/T_1$ shows thermally activated behavior below $T_{\rm P}$, as expected for the PS phase. To account for the observed $1/T_1 \propto T^{0.6}$ at the QPT, we use the relation $1/T_1 = A \, T^{0.6} e^{-\Delta / k_{\rm B}T}$, with $A$ a field-dependent constant, as indicated by the solid lines in Fig.~\ref{relax}(a). The energy gaps, $\Delta$, extracted from these fits are described by the linear function $\Delta = g \mu_B (H_{\rm C} - H)$ shown in Fig.~\ref{pd}, with a $g$-factor of 2.28~\cite{kageyama_JPSJ_1998} and a critical field $H_{\rm C} \simeq 5.5$~T. By linear extrapolation, we obtain $\Delta \approx 8.3$~K at zero field, consistent with the 8.5~K gap reported from specific-heat measurements~\cite{Guo_2023}.

For fields of 6~T and above, an onset temperature $T^*$ remains present, also marked by an upturn in $1/T_1$ with decreasing $T$ [Fig.~\ref{relax}(b)]. The N\'{e}el temperature, $T_{\rm N}$, is identified by the sharp drop in $1/T_1$, marked by the arrows in the figure, below which the data show gapless behavior; $T_{\rm N}$ increases monotonically with field to 1.5~K at 10~T. The broad peak appearing in $1/T_1$ between $T_{\rm N}$ and $T^*$ for all fields above 6~T suggests the emergence of quasi-static AFM order, which we denote as an ``antiferromagnetic liquid'' (AFL) regime.

{\it Deconfined quantum criticality---}The values of $T^*$, $T_{\rm P}$, and $T_{\rm N}$ we extract from P$_1$ allow us to construct the $(H, T)$ phase diagram at 2.4~GPa shown in Fig.~\ref{pd}. The results reveal field-induced suppression of the EP phase and emergence of the AFM phase, with the QPT occurring near 5.5~T. The gap-closing field determined from $1/T_1$ aligns with the disappearance of $T_{\rm P}$ deduced from $K_{\rm n}$. By comparison, the phase diagram at the same pressure based on P$_2$~\cite{Cui_Science_2023} shows a very similar critical field (5.7~T) despite the markedly different forms of $K_{\rm n}(T)$ and $1/T_1(T)$ on P$_1$ and P$_2$ [Figs.~\ref{spec}(b,c)]. With both ordering temperatures suppressed to such a low value at 5.5~T, the field-induced QPT in P$_1$ appears to be continuous or nearly continuous. Because the two phases have unrelated symmetries, this is the primary hallmark of a DQCP. We remark for context that
transitions between gapped valence-bond states and ordered magnetic states are known in other condensed-matter systems, notably insulating organic charge-transfer salts~\cite{Kanoda_2011}, but in all known cases the transition is thermal~\cite{Miksch_Science_2021}, or the QPT appears to be first-order~\cite{Shimizu_PRR_2021}, and there is nowhere in phase space at which the energy scales of both unrelated phases vanish together as required at a DQCP.

The enlarged phase space we have discovered contains a multicritical point where the FP, EP, and AFM phases meet [Fig.~\ref{pd2d}(a)]. Some authors have already considered this scenario using different models \cite{Lee_PRX_2019,Xi_PRB_2023}, finding an emergent enhanced SO(5) symmetry of the ``parent'' DQCP at the multicritical point, with the phase boundaries forming lines of more ``generic'' O(4) DQCPs, or of weakly first-order QPTs with O(3) symmetry~\cite{Cui_Science_2023}. Other authors have discussed the presence of an extended quantum spin-liquid phase adjoining the multicritical DQCP~\cite{Yang_PRB_2022, Liu_PRX_2022}. Some light can be shed on the DQCP symmetry by considering the exponent, or anomalous dimension, $\eta \approx 0.6$, extracted from $1/T_1$ at P$_1$ in the quantum critical regime~\cite{Senthil_PRB_2004, LiCC_JPCM_2024}, whose relation to the CFT scaling dimension $\Delta_1$ is consistent with an SO(5) symmetry. By contrast, the smaller value $\eta \approx 0.2$ observed for P$_2$~\cite{Cui_Science_2023} is consistent with the symmetry of a generic DQCP. However, the wide range of $\eta$ values observed by different numerical methods for models with different emergent symmetries ~\cite{Sandvik_PRL_2007, Nahum_PRX_2015, Mengzy_PRX_2017, Guzc_PRL_2024} complicates a direct determination of the DQCP universality by comparison of experiment with theory.

{\it Discussion---}The primary issue in unifying our NMR results for SrCu$_2$(BO$_3$)$_2$ with theoretical DQCP scenarios is the role of the magnetic field. In Fig.~\ref{pd2d}(b) we outline the phase diagram of SrCu$_2$(BO$_3$)$_2$ based on our experimental control of both pressure and field. The FP and EP coexist over a range of pressures due to the first-order phase transition. The application of field is required to induce proximate DQCP behavior at both the FP--AFM and the EP--AFM transitions; the emergent symmetries on these lines of QPTs are thought to be O(3) and O(4) respectively, and so we mark the parent SO(5) point where the two lines meet. Missing from the DQCP theories are the weak Dzyaloshinskii-Moriya interactions of SrCu$_2$(BO$_3$)$_2$ \cite{Cepas_PRL_2001}, which even at high pressure are thought to be weaker than the DQCP field scales of order 5~T that we find. The open issue is whether the field is simply closing small gaps (erasing small first-order character) induced by these terms, revealing the anticipated zero-field DQCP behavior with no significant changes, or whether it does affect the competition of $J'$ and $J$ terms in the PS and AFM phases.

Our results show that the field-induced changes in $T_{\rm P}$ and $T_{\rm N}$ are much richer than the effects observed under pressure at zero field~\cite{Guo_2023}. For theory, this highlights the importance of investigating whether a multicritical DQCP~\cite{You_PRB_2021, Yang_PRB_2022, Xi_PRB_2023} or a quantum spin liquid~\cite{Yang_PRB_2022} can be stabilized at finite fields. On the experimental side, the pressure-induced PS--AFM transition was achieved in a diamond anvil cell~\cite{Guo_2023}, and further studies ensuring highly hydrostatic conditions are required to revisit the zero-field case.

A remaining question is how to understand the role of macroscopic phase coexistence, arising from the first-order FP--EP transition, on the DQCP. At least the mechanism of this transition is consistent with a lattice effect~\cite{Boos_PRB_2019}, albeit one that is both spontaneous and relative (in the sense that a constant spin-lattice coupling can no longer drive a distortion when the lattice is sufficiently constrained at high pressures). Concerning the energetics and fluctuations of the two PS states, the inclusion of a dimer in the FP singlet (exclusion in the EP) implies that the DS phase competes more strongly with the EP phase, whereas the FP phase shares some of its correlations. This in turn
leads to the opposing spin-fluctuation effects we observed in the P$_1$ (EP) and P$_2$ (FP) signals, where the latter become gapped in the
PL regime whereas the former gap only when the PS is established.

{\it Summary---}Our NMR study of the Shastry-Sutherland compound SrCu$_2$(BO$_3$)$_2$ at pressures up to 2.65~GPa reveals an empty-plaquette (EP) phase coexisting with the previously observed full-plaquette (FP) phase. The quite different spin dynamics of the two peaks we observe allow us to associate them with these two distinct phases. Their macroscopic coexistence and systematic changes in volume fraction indicate a first-order pressure-induced FP-to-EP transition. Focusing on the EP phase, we observe a field-induced EP--AFM phase transition near 5.5~T, accompanied by the power-law scaling behavior, $1/T_1 \propto T^{0.6}$ in the quantum critical regime. We propose that the extra dimension of the quantum phase space is controlled by a spin-lattice coupling. This additional dimension has direct consequences for the nature and symmetry of the DQCP, specifically pointing to a multicritical point with emergent SO(5) symmetry. Our experimental realization of the EP phase of the SSM through combined pressure and field control points the way to a systematic mapping of the high-dimensional phase space and complex landscape of deconfined quantum criticality in SrCu$_2$(BO$_3$)$_2$.

{\it Acknowledgments---}We thank W. Li for helpful discussions. This work was supported by the National Key Research and Development Program of China (Grant No.~2023YFA1406500), the Scientific Research Innovation Capability Support Project for Young Faculty (Grant No.~ZYGXQNJSKYCXNLZCXM-M26), and the National Natural Science Foundation of China (Grant Nos.~12374156, 12134020, 12334008, and 12174441). A portion of this work was conducted at the Synergetic Extreme Conditions User Facility (SECUF).


%

\end{document}